\documentclass[aps,prl,twocolumn,showpacs]{revtex4}
\usepackage{amsmath,amssymb,graphicx}
\begin{document}
\newcommand{\T}{{\mathcal T}}
\title{\large Mean escape time over a fluctuating barrier}
\author{Jan Iwaniszewski}
\email{jiwanisz@phys.uni.torun.pl}
\affiliation{Institute of Physics, Nicolaus Copernicus University,
Grudzi\c{a}dzka~5, 87-100 Toru\'n, Poland }
\received{\today}
\begin{abstract}
An approximate method for studying activation over a fluctuating barrier of potential is proposed. It involves considering separately the slow and fast components of barrier fluctuations, and it applies for any value of their correlation time $\tau$. It gives exact results for the limiting values $\tau \rightarrow 0$ and $\tau \rightarrow \infty$, and the agreement with numerics in between is also excellent, both for dichotomic and Gaussian barrier perturbations.
\end{abstract}
\maketitle

%%%%%%%%%%%%%%%%%%%%%%%%%%%%%%%%%%%%%%%%%%%%%%%%%%%%%%%%%%%%%%%%%%%%%

\par
Ever since Kramers seminal paper \cite{kra40} the fluctuational escape over a potential barrier has been a paradigm for a thermal activation process. Recently, activation in the presence of time-varying fields have become a subject of great interest due to the discovery of many counterintuitive noise-assisted effects, like stochastic resonance \cite{gam98} or transport in Brownian motors \cite{rei02}. The nonequilibrium character of these problems hinders, however, the direct application of many ideas and methods developed for investigation of the static Kramers problem \cite{han90} (e.g., detailed balance or rate concept). On the other hand, as the time-scale of variation of the driving signal is independent of the internal dynamics of the system, standard adiabatic methods are restricted to certain ranges of parameters, only. Hence, an approach which overcomes these difficulties and applies for the whole range of time variability of the perturbation, is of great importance. 

\par
In this letter we address this problem for an activation over a randomly fluctuating barrier. The subject is interesting not only due to its ubiquity in many branches of physics, e.g., in relation to ligand binding to heme proteins \cite{agm83b}, transport processes in glasses \cite{bin84}, or dye laser with a fluctuating pump parameter \cite{kam81}, but especially because of the phenomenon of {\it resonant activation} \cite{doe92} --- the appearance of a minimum of the mean activation time $\T$ as a function of the correlation time $\tau$ of barrier fluctuations. The dependence $\T(\tau)$ can be calculated exactly merely for simple models \cite{doe92,van93,zurbre}, for more general cases the approaches \cite{rei95b,rei96,rei98,madank,biebie} proposed till now apply to some ranges of $\tau$, only. Irrespective of technical differences they are all  based on the rate concept, which assumes a quasi-stationary equilibrium before the activation happens and applies for $\tau\ll \T$, and/or kinetic description for $\tau \gg \ln(1/q)$ ($q$ states for the thermal noise intensity) when the escape events are uncorrelated with the potential variations. Although for small enough $q$ in an extended region $\ln(1/q) \ll \tau \ll \T$ both approximations coexist and give similar results \cite{rei95b}, nevertheless the proper smooth connection between them remains the main theoretical challenge. Below, we present an approach which is valid for any $\tau$. It gives exact values of $\T(\tau)$ in the limits $\tau \rightarrow 0$ and $\tau \rightarrow \infty$, and a very good approximation in between. 

\par
We study an overdamped Brownian particle driven by a (thermal) Gaussian white noise $\xi(t)$ of zero mean, which moves in a stochastically varying potential. Its static part $U(x)$ has a monostable or bistable form and the random part $V(x)z(t)$ is generated by a stationary Markovian noise $z(t)$ of zero mean and correlation $C(t)=Q/\tau \exp(-|t|/\tau)$. Following \cite{iwa96,iwa00a} we assume a general form for its intensity $Q(\tau) =  Q_0 \tau^{\alpha}$ ($0 \le Q_0=\mathrm{const} \, , \; 0\leq\alpha\leq1$), which gives the mostly studied cases with $\tau$-independent intensity ($\alpha=0$) or variance ($\alpha=1$) as special cases. Two types of $z(t)$ are considered: an Ornstein-Uhlenbeck noise (OUN), which is Gaussian with variance $D=Q/\tau$, and a dichotomic noise (DN), which flips between two values $\pm\sqrt{D}$ with the rate $\gamma=1/(2\tau)$. Although they essentially differ --- the former is continuous, the latter discrete --- nevertheless, they influence the activation process very similarly \cite{mar96} and the main steps of the presented description are the same. The dynamics of the system is given by the nonmarkovian Langevin equation
\begin{equation}
\label{langevin} 
\frac{dx}{dt} = -\, U'(x) -  V'(x)z(t) + \xi(t) \,  .
\end{equation}
Introducing the two-dimensional markovian stochastic process $\{x(t),z(t)\}$ one can formulate the evolution equation for the joint probability distribution $P(x,z,t)$:
\begin{equation} 
\label{fpe} 
\frac{\partial}{\partial t} P(x,z,t) = 
\left[{\mathcal L}(x,z) + \Lambda(z) \right] P(x,z,t) \, , 
\end{equation}
where ${\mathcal L}(x,z)= \frac{\partial}{\partial x} \left[U'(x)+V'(x)z\right] +q\frac{\partial^2}{\partial x^2}$ is the Fokker-Planck (FP) operator. The free evolution of the barrier noise is governed by the operator $\Lambda(z)=1/{\tau}\left[({\partial}/{\partial z})z + Q{\partial^2}/{\partial z^2}\right]$ for OUN or by the matrix $\Lambda  =  (-\gamma,\gamma;\gamma,-\gamma)$ for DN. Initially the particle is located at the bottom $x_b$ of the well and the quantity of interest is the mean first passage time (MFPT) through a given threshold $x_{thr}$ located either at the top $x_t$ or far from it on the other side of the barrier. 

%%%%%%%%%%%%%%%%%%%%%%%%%%%%%%%%%%%%%%%%%%%%%%%%%%%%%%%%%%%%%%%%%%%%%

\par
A typical scenario of an escape event consists of two stages. For a long time $t_b$ the particle fluctuates in the vicinity of the bottom of the well, being subjected to small random impacts of $\xi(t)$. If a large enough outburst of $\xi(t)$ occurs the particle will eventually surmount the barrier almost immediately, during a short time $t_t$. The time-variation of the potential exerts only a negligible effect on the first stage, but it can essentially modify the dynamics during the second one, when the particle interacts with the whole slope of the barrier. Any realization of $\xi(t)$, which has been supposed to bring the particle over the top of a static barrier, may turn out to be insufficient if the barrier rises during the climbing stage. On the contrary, if the barrier decreases the particle does cross to the other side, but some smaller outbursts of $\xi(t)$ would also result in a successful escape. Because the rate of variation of the barrier shape depends on the correlation time of $z(t)$, the relationship between $t_t$ and $\tau$ appears to be crucial in the analysis \cite{iwa00a}.

\par 
This discussion leads us to the central idea of the present approach -- splitting the barrier noise into two independent components: 
\begin{equation} 
z(t)=z_s(t)+z_f(t) \; . 
\label{zsf} 
\end{equation}
The slow one $z_s$ is defined as the mean value of $z$ over the time interval of climbing $(t,t+t_t)$ and over its possible realizations (marked by $\langle...\rangle$)
\begin{equation} 
\label{zs} 
z_s(t)=\left<\frac{1}{t_t} \int_t^{t+t_t} ds \, z(s) \right> 
= \frac{1}{\Delta} \left(1-e^{-\Delta}\right) z(t) \, , 
\end{equation} 
where $\Delta = t_t/\tau$. It is supposed to be constant during the climbing stage, while its random character arises from the randomness of $z(t)$. Hence $z_s$ is governed by the same statistics as $z$ but with the variance
\begin{equation} 
\label{ds} 
D_s=\left< z_s^2 \right> = \frac{Q}{t_t} \, \frac{1}{\Delta}\left(1-e^{-\Delta}\right)^2 \, . 
\end{equation} 
Next, assuming that fast part $z_f(t)$, which gives rapid fluctuations around $z_s(t)$, can be treated as uncorrelated, one calculates its intensity $Q_f$:
\begin{equation} 
\label{qf} 
Q_f=Q \left[ 1-\frac{1}{\Delta}\left(1-e^{-\Delta}\right) 
-\frac{1}{2}\frac{1}{\Delta}\left(1-e^{-\Delta}\right)^2 \right] \, . 
\end{equation} 
If $z(t)$ is Gaussian it can always be written as the sum of two independent Gaussian components (\ref{zsf}). So, in OUN case both $z_s(t)$ and $z_f(t)$ are OUN's with correlation time $\tau$ and they differ only in the form of their intensities (variances) $D_i=Q_i/\tau$ ($i=f,s$). If $\tau \rightarrow 0$ one has $Q_f = Q$, while the leading-order term of $Q_s$ reads $Q/\Delta^2$ so, for any $\alpha$, it vanishes at least linearly with $\tau$. Thus one is left with only the fast part of $z(t)$. In the opposite limit $\tau \rightarrow \infty$ the leading term of $Q_f$ becomes $Q\Delta^2/3$, so $D_f$ vanishes at least linearly with $1/\tau$, while $D_s=D$. Only the slow part of $z(t)$ survives. One can check that, ignoring the dependence of $Q$ on $\tau$, the intensities $Q_f$ and $Q_s$ are monotonic functions of $\tau$. While for $\tau=0$ one has the white-noise limit of $z(t)$ with rapid fluctuations $z_f(t)$, an increase of $\tau$ increases the role of $z_s$ at the expense of decrease of the intensity of $z_f$, eliminating it completely as $\tau \rightarrow \infty$. Thus the fundamental difference between $z_s(t)$ and $z_f(t)$ consists in the different regime of values of $\tau$ in which they exist: $z_s(t)$ occurs for $\tau \gtrsim t_t$, and hence fluctuates slowly, while $z_f(t)$ persists for $\tau \lesssim t_t$ and varies rapidly. Only for $\tau\sim t_t$ do they coexist.  

\par
A similar summation property to that for Gaussian noise does not apply to the dichotomic noise --- one cannot display a given dichotomic noise as the sum of two independent dichotomic noises. However, the great similarity between the statistical properties of OUN and DN suggests treating the DN case in the same way. The definitions (\ref{zsf}) and (\ref{zs}) involve the asymmetric character of two-state noise $z_f(t)$ and its dependence on $z_s(t)$, but for simplicity we assume, that both $z_s(t)$ and $z_f(t)$ are symmetric, independent dichotomic noises of zero mean. Since OUN and DN have the same correlation function the formulas (\ref{zs}-\ref{qf}) apply to the DN case, as well.  

\par
\begin{figure}[t]
\centering
\includegraphics[width=0.88\columnwidth]{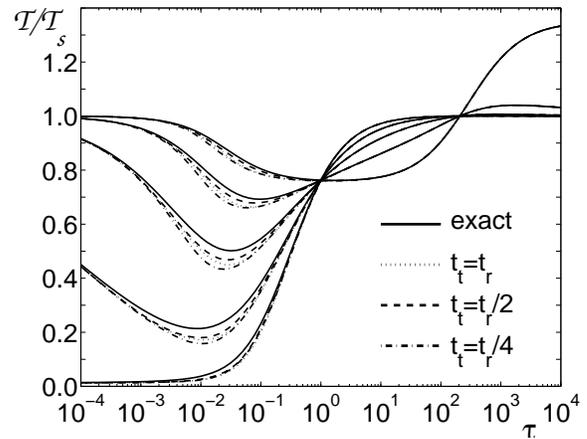}
\caption{Relative mean escape time ${\T / \T_s}$ versus $\tau$ for DN case with a triangle barrier $U(x)=10(1-|x|)$ and $V(x)=1-|x|$ confined to the interval $(-1,1)$ for $Q_0=1$, $q=1$, $x_{thr}=0$, $t_r=0.09$ and  $\alpha=0,\,0.25,\,0.50,\,0.75,$ and $1.0$ from the bottom to the top on the left-hand-side, respectively. Solid lines show the exact results and the others the approximation ({\ref{mfpt}}) for few values of $t_t$.}
\label{fdg}
\end{figure}

 
We should also determine the value of integration interval $t_t$. For an unperturbed potential it equals the relaxation time $t_r$ from the top to the bottom of the well, but fluctuations of the potential lead to far-from-equilibrium conditions, so that this equality does not hold \cite{mailuc}. However, we do not intend here to consider the relationship between the processes of climbing up and relaxing down the fluctuating barrier. Rather, we need a tool for calculating the order of the duration of the second stage of the escape event. It is enough to take for it the value of $t_r$ for a static barrier, which may be calculated as the MFPT from the top $x_t$ to the bottom $x_b$ of the well. It is shown in Fig. \ref{fdg} that our results depend almost unnoticeably on the variation of $t_t$ within the range of tens of percent. A more careful analysis would require us to take into account, not only the mean value, but also the statistical distribution of relaxation times \cite{biebie}.

%%%%%%%%%%%%%%%%%%%%%%%%%%%%%%%%%%%%%%%%%%%%%%%%%%%%%%%%%%%%%%%%%%%%%

\par 
Using the decomposition (\ref{zsf}) the escape problem may be considered as a three-dimensional markovian process. Its joint probability distribution $P(x,z_f,z_s,t)$ evolves accordingly to the FP equation similar to (\ref{fpe}) but with two $\Lambda$'s operators for $z_f$ and $z_s$ (with $Q_f$ or $Q_s$ instead of $Q$, respectively), and $z=z_f+z_s$ in ${\mathcal L}(x,z)$. Such a formulation allows for a clear separation of different time-scales of the system dynamics. Since, by definition, $z_s$ remains constant while the particle climbs the barrier, its dynamics may be analyzed by the  kinetic approach. On the contrary, $z_f$ vanishes for $\tau$ slightly greater than $t_r$, but still for $\tau\ll\T$, so rate theory applies. Thus we seek the probability distribution in the form $P(x,z_f,z_s,t)=p(x,z_f,t;z_s)\rho(z_s,t)$ \cite{ber98b}. 

\par
The fast equilibration process is described by the evolution of $p(x,z_f,t;z_s)$, which is governed by the equation
\begin{equation} 
\label{rfpe} 
\frac{\partial}{\partial t} p(x,z_f,t;z_s) = 
\left[{\mathcal L}(x,z_f;z_s) + \Lambda(z_f) \right] p(x,z_f,t;z_s) \, , 
\end{equation}
where ${\mathcal L}(x,z_f;z_s)= \frac{\partial}{\partial x} \left[{\mathcal U}'(x;z_s)+V'(x)z_f\right] +q\frac{\partial^2}{\partial x^2}$ and the slow component of barrier fluctuations gives rise to different forms of potential configurations ${\mathcal U}(x;z_s)=U(x)+V(x)z_s$. Following a standard method one looks for the quasi-potential $\Phi$ being the dominant exponential term of the reduced (quasi-)stationary probability distribution $\left\langle p(x,z_f;z_s)\right\rangle_{z_f} = p(x;z_s) \sim \exp\left[-\Phi(x;z_s)/q\right]$ of (\ref{rfpe}). For the DN case we obtain an equation   
% 
%\begin{widetext}
%\begin{equation} 
%\label{phi} 
%\Phi'(x;z_s)\left[{\mathcal U}'(x;z_s)-\Phi'(x;z_s)\right]^2 = \frac{q}{\tau} %\left[G(x)\Phi'(x;z_s) - {\mathcal U}'(x;z_s)\right] \;, 
%\end{equation}
%\end{widetext}
%
% 
\begin{eqnarray}
\label{phi} 
\Phi'(x;z_s)&&\hspace{-10pt} \left[{\mathcal U}'(x;z_s)-\Phi'(x;z_s)\right]^2 \nonumber \\
=&& \hspace{-10pt}\frac{q}{\tau} \left[G(x)\Phi'(x;z_s) - {\mathcal U}'(x;z_s)\right] \;, 
\end{eqnarray}
whose middle (of the three always real) solution gives the quasi-potential. This equation is formally similar to the result of Reimann and Elston \cite{rei96}, who consider the case $\tau\ll\T$, however. The only difference is the form of diffusion function $G(x)=1+(Q_f/q) V'(x)^2$. In \cite{rei96} the total noise intensity $Q$ is used, what gives an improper limiting value of $\Phi'$ for $\tau \rightarrow \infty$ for $\alpha=1$. Here $G(x)$ depends on $Q_f$, which vanishes for any $\alpha$ as $\tau\rightarrow\infty$, so one obtains the exact expression $\Phi'(x;z_s)\rightarrow {\mathcal U}'(x;z_s)$. In the opposite limit of $\tau\rightarrow 0$, the solution of (\ref{phi}) converges to the exact form $U'(x)/G(x)$. This suggests to deal not with the quasi-potential but rather with an effective one
\begin{equation} 
\label{effu} 
{\mathcal {\mathcal U}}_{eff}'(x;z_s) = \Phi'(x;z_s) G(x) \; .
\end{equation} 
Finally, exploiting the well known form of the exact FP equation in the white-noise-limit \cite{fox86}, one can write the effective FP operator
\begin{equation} 
\label{efpe} 
{\mathcal L}_{eff}(x;z_s) 
\equiv \frac{\partial}{\partial x} {\mathcal {\mathcal U}}_{eff}'(x;z_s)\, + \,q\frac{\partial}{\partial x}\sqrt{G(x)} 
\frac{\partial}{\partial x}\sqrt{G(x)} \,,
\end{equation} 
which governs the fast part of the evolution.

\par
A convenient way of finding the quasi-potential in the OUN case formulates the problem by means of path-integral or Hamiltonian techniques \cite{rat91}. In general, the problem cannot be elaborated analytically, but asymptotic expressions for small and large $\tau$ are available \cite{rei95b,rei98}. To attempt an interpolation between the two limits of $\tau$ we construct a 2-2 Pad\'{e} approximant \cite{bak86}
%
%\begin{widetext}
%\begin{equation}
%\label{pade}
%\Phi'(x;z_s) = \frac{{\mathcal U}'(x;z_s)}{G(x)} 
% \frac{1 + \tau\, 2Q_f V'(x)^2{\mathcal W}(x;z_s)/G(x) + {\tau}^2 \, {\mathcal %W}(x;z_s)^2 } {1 + {\tau}^2\, {\mathcal W}(x;z_s)^2/G(x)} \,, 
%\end{equation}
%\end{widetext}
%
%
\begin{eqnarray}
\label{pade}
 &&\hspace{-8mm} \Phi'(x;z_s) = \frac{{\mathcal U}'(x;z_s)}{G(x)}\times \\
&& \hspace{-5mm}\times\frac{1 + \tau\, 2Q_f V'(x)^2{\mathcal W}(x;z_s)/G(x) + {\tau}^2 \, {\mathcal W}(x;z_s)^2 } {1 + {\tau}^2\, {\mathcal W}(x;z_s)^2/G(x)} \,, \nonumber
\end{eqnarray}
with ${\mathcal W}(x;z_s)=\left[{G(x)}/{V'(x)}\right]\left[{\mathcal U}\,'(x;z_s)V'(x)/G(x)\right]'$. One can notice, that as a function of $\tau$ the expression (\ref{pade}) has no singularities and monotonically increases with $\tau$, what is an anticipated property of quasi-potential \cite{rat91,rei98}. As for DN, we may also introduce an effective potential (\ref{effu}). Using (\ref{efpe}) calculation of the MFPT $\T(z_s)$ for both types of barrier noise is straightforward. 

%%%%%%%%%%%%%%%%%%%%%%%%%%%%%%%%%%%%%%%%%%%%%%%%%%%%%%%%%%%%%%%%%%%%%
 
\par
In the slow time scale the evolution of the system is governed by the Smoluchowski equation with a sink term 
\begin{equation}
\label{smols}
\frac{\partial}{\partial t}\rho(z_s,t) \, = \, \left[\Lambda(z_s) -
k(z_s)\right] \rho(z_s,t) \; .
\end{equation}
It describes stochastic switchings between the potential configurations of different $z_s$ and an escape process from each of them [$k(z_s)=\mu/\T(z_s)$ with $\mu=1/2$ for $x_{thr}=x_t$, or $\mu=1$ for $x_{thr}$ far from it]. One gets the mean escape time integrating $\rho(z_s,t)$ over $t\in(0,\infty)$, and summing/integrating over $z_s$ for DN/OUN. For the dichotomic switching the result is immediate:
\begin{equation}
\label{mfpt}
\T=\frac{2\T_+\T_-+ \mu\,\tau(\T_++\T_-)}{\T_++\T_-+2\mu\,\tau} \; ,
\end{equation}
where $\T_{\pm}$ are the MFPT's for ${\mathcal U}_{\pm}(x)=U(x)\pm\sqrt{D_s}V(x)$, respectively. Although (\ref{mfpt}) resembles the well-known solution \cite{van93,bie93}
of a very simple set of equations (\ref{smols}) which constitutes the long-$\tau$ approximation of the problem \cite{rei96}, the dependence of $\T_+$ and $\T_-$ on $Q_f(\tau)$ involves also the fast part of the dynamics in the formula (\ref{mfpt}).

\par
The problem is much more complicated in the OUN case. To the best of the author's knowledge there is no universal approximation of (\ref{smols}) valid for any $\tau$ \cite{ber98c}. One may calculate asymptotic expressions for small and large $\tau$ \cite{agmrab,rei95b} and construct a Pad\'{e} approximant to interpolate in between; however, the complicated exponential dependence of expansion terms on the amplitude of fluctuations yields a very bad approximation \cite{iwa03}. Hence, in what follows, we solve (\ref{smols}) numerically. 

%%%%%%%%%%%%%%%%%%%%%%%%%%%%%%%%%%%%%%%%%%%%%%%%%%%%%%%%%%%%%%%%%%%%%

%%%%%%%%%%%%%%%%%%%%%%%%%%%%%%%%%%%%%%%%%%%%%%%%%%%%%%%%%%%%%%%%%%%%%

\par
\begin{figure}[t]
\centering
\includegraphics[width=0.88\columnwidth]{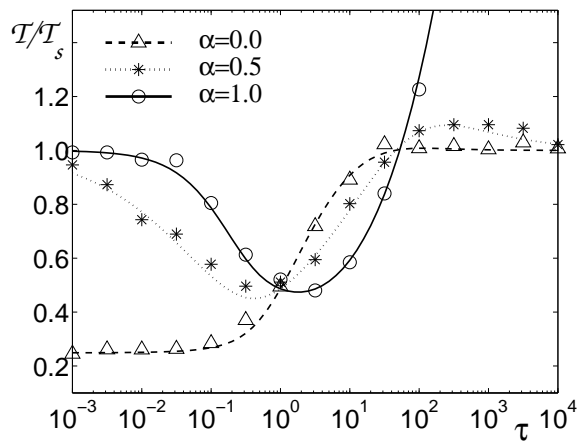}
\caption{Relative mean escape time ${\T / \T_s}$ versus $\tau$ for OUN case and the system with $U(x)=x^4/4-x^2/2$, $V(x) = U(x)+1/4$ for $|x|\leq 1$ and $V(x)=0$ elsewhere, for $q=0.08$, $Q_0=0.8$, $t_t=t_r/2=1.26$, and three values of $\alpha$. The lines present our approximation and markers are from the numerical simulation of (\ref{langevin}).}
\label{foun}
\end{figure}

To test the method we take the triangular barrier model \cite{doe92} with DN. In Fig. \ref{fdg} we plot $\T(\tau)/\T_s$ ($\T_s$ is the MFPT for a static barrier) for the exact analytical results and for the present method, in each case for few values of $\alpha$. The relaxation time calculated from the exact formula \cite{biebie} for the MFPT from $x_t=0$ to $x_b=1$ equals $t_r=0.09$. We show three sets of curves with $t_t=t_r$, $t_t=t_r/2$ and $t_t=t_r/4$, respectively. The agreement with the exact plot is very good, but in the interval $10^{-3}<\tau<10^{-1}$ our method gives slightly lower values. We have found the smallest deviation for $t_t=t_r/2$, but even when $t_t$ is twice larger or smaller the difference is still not very significant. This validates the way we estimate the interval of integration $t_t$ in (\ref{zs}). For simplicity in the next example we use $t_t=t_r/2$, but to be more precise, for each system a careful analysis of its best value should be done \cite{iwa03}. In Fig. \ref{foun} we display $\T(\tau)/\T_s$ for OUN case and three values of $\alpha$. The agreement between the theory and numerical simulation of (\ref{langevin}) is very good, but also with some underestimation in the region of the resonant activation minima. The results for other systems and other values of parameters are also excellent \cite{iwa03}.

%%%%%%%%%%%%%%%%%%%%%%%%%%%%%%%%%%%%%%%%%%%%%%%%%%%%%%%%%%%%%%%%%%%%%

%%%%%%%%%%%%%%%%%%%%%%%%%%%%%%%%%%%%%%%%%%%%%%%%%%%%%%%%%%%%%%%%%%%%%

\par
To conclude, we have presented a method of an investigation of thermal activation in the presence of barrier fluctuations for arbitrary duration of their correlation. Dividing the barrier noise into two components -- the slow and fast ones -- we can separate two time scales of the evolution of the system for any value of $\tau$ and use both rate and kinetic approaches in the analysis without any sewing procedure. The noise division is done through an averaging over a finite interval of time $t_t$ (\ref{zs}), hence we call the approach {\it a partial noise-averaging method} (PNAM). For a dichotomic perturbation the formula (\ref{mfpt}) together with the MFPT obtained for the FP operator (\ref{efpe}) provides for the first time the analytical expression for the dependence ${\mathcal T}(\tau)$ for any $\tau\in [0,\infty)$, for arbitrary potentials $U(x)$ and $V(x)$, and a large class of noises. For the OUN we have been obliged to use a computer at the final step, but the accordance of the present result with the full-numerical ones confirms the power of PNAM. Although the method is presented in terms of MFPT, it can be expressed by means of any of the standard approaches \cite{han90} to the activation process. We hope also, that the presented idea of splitting the noise could be useful in other problems where different time-scales coexist, making the proposed approach valuable for many applications. 

%%%%%%%%%%%%%%%%%%%%%%%%%%%%%%%%%%%%%%%%%%%%%%%%%%%%%%%%%%%%%%%%%%%%%

\par
The research was partially supported by the Royal Society, London. The author is very indebted to Prof. P.~V.~E.~McClintock for his kind hospitality in Lancaster where the basic ideas of the present work were born.

%%%%%%%%%%%%%%%%%%%%%%%%%%%%%%%%%%%%%%%%%%%%%%%%%%%%%%%%%%%%%%%%%%%%%

%%%%%%%%%%%%%%%%%%%%%%%%%%%%%%%%%%%%%%%%%%%%%%%%%%%%%%%%%%%%%%%%%%%%%


\begin{thebibliography}{30}
\bibitem{kra40}
H.A.~Kramers, Physica {\bf 7}, 284 (1940).
\bibitem{gam98}
L.~Gammaitoni \textit{et al.}, Rev. Mod. Phys. \textbf{70}, 223 (1998).
\bibitem{rei02}
P.~Reimann, Phys. Rep. \textbf{361}, 57 (2002).
\bibitem{han90}
P.~H\"anggi, P.~Talkner, and M.~Borkovec, Rev. Mod. Phys. \textbf{62}, 251 (1990).
\bibitem{agm83b}
N.~Agmon and J.J.~Hopfield, J.~Chem. Phys. \textbf{79}, 2042 (1983).
\bibitem{bin84}
K.~Binder and A.P.~Young, Rev. Mod. Phys. \textbf{58}, 801 (1986).
\bibitem{kam81}
K.~Kaminishi \textit{et~al.}, Phys. Rev. \textbf{24},
370 (1981).
\bibitem{doe92}
C.R.~Doering and J.C.~Gadoua, Phys. Rev. Lett.  \textbf{69}, 2318 (1992).
\bibitem{van93}
C. Van den Broeck, Phys. Rev. E \textbf{47},
4579 (1993).
\bibitem{zurbre}
U.Z\"urcher and C.R.~Doering, Phys. Rev. E  \textbf{47}, 3862 (1993); J.J.~Brey and J.~Casado-Pascual, {\itshape{ibid}} \textbf{50}, 116 (1994).
\bibitem{rei95b}
P.~Reimann, Phys. Rev. E  \textbf{52}, 1579 (1995).
\bibitem{rei96}
P.~Reimann and T.C.~Elston, Phys. Rev. Lett.  \textbf{77}, 5328 (1996).
\bibitem{rei98}
P.~Reimann, R.~Bartussek, and P.~H\"anggi, Chem. Phys.  \textbf{235}, 11 (1998).
\bibitem{madank}
A.J.R. Madureira \textit{et~al.}, Phys. Rev. E  \textbf{51}, 3849 (1995);
J.~Ankerhold and P.~Pechukas, Physica A \textbf{261}, 458 (1998).
\bibitem{biebie}
M.~Bier and R.D.~Astumian, Phys. Lett. A  \textbf{247}, 385 (1998);
M.~Bier \textit{et~al.}, Phys. Rev. E  \textbf{59}, 6422 (1999).
\bibitem{iwa96}
J.~Iwaniszewski, Phys. Rev. E  \textbf{54},
3173 (1996).
\bibitem{iwa00a}
J.~Iwaniszewski \textit{et~al.}, Phys. Rev. E  \textbf{61}, 1170 (2000).
\bibitem{mar96}
M.~Marchi \textit{et~al.}, Phys. Rev. E  \textbf{54},
3479 (1996).
\bibitem{mailuc}
R.S.~Maier and D.L.~Stein, Phys. Rev. E  \textbf{48}, 931 (1993); D.G.~Luchinsky and P.V.E.~McClintock, Nature \textbf{389}, 463 (1997).
\bibitem{der99}
I.~Der\'enyi and R.D.~Astumian, Phys. Rev. Lett. \textbf{82}, 2623 (1999).
\bibitem{ber98b}
A.M.~Berezhkovskii \textit{et~al.}, Physica A \textbf{251}, 399 (1998).
\bibitem{fox86}
R.F.~Fox, Phys. Rev. A \textbf{33}, 467 (1986).
\bibitem{rat91}
K.M.~Rattray and A.J.~McKane, J.~Phys. A \textbf{24},
1215 (1991).
\bibitem{bak86}
G.A.~{Baker,~Jr.} and P.~Graves-Morris,   \emph{Pad\'{e} approximants} (Addison-Wesley, London,
1981).
\bibitem{bie93}
M.~Bier and R.D.~Astumian, Phys. Rev. Lett. \textbf{71},
1649 (1993).
\bibitem{ber98c}
A.M.~Berezhkovskii, Y.A.~D'yakov, and V.Y.~Zitserman, J.~Chem. Phys. \textbf{109},
4182 (1998).
\bibitem{agmrab}
N.~Agmon, J.~Chem. Phys. \textbf{90}, 3765 (1989); 
S.~Rabinovitch and N.~Agmon, Chem. Phys. \textbf{148},
11 (1990).
\bibitem{iwa03}
J.~Iwaniszewski, to be published.

\end{thebibliography}
\end{document}